\documentclass[%
 reprint,
 amsmath,amssymb,
 aps,
 prappl,
]{revtex4-2}

\usepackage{graphicx}
\usepackage{dcolumn}
\usepackage{bm}

\newcommand\days{200}
\usepackage{gensymb} 

\begin{document}

\preprint{APS/123-QED}

\title{A passively pumped vacuum package sustaining cold atoms for more than \days\ days}

%
%
%
%
\author{Bethany J. Little}
\email{bjlittl@sandia.gov}

\author{Gregory W. Hoth}

\author{Justin Christensen}

\author{Chuck Walker}

\author{Dennis J. De Smet}
\affiliation{Sandia National Laboratories, Albuquerque, NM 87185, USA}

\author{Grant W. Biedermann }
\affiliation{Department of Physics and Astronomy, University of Oklahoma, Norman, Oklahoma 73019, USA}

\author{Jongmin Lee }
\affiliation{Sandia National Laboratories, Albuquerque, NM 87185, USA}

\author{Peter D. D. Schwindt}
\affiliation{Sandia National Laboratories, Albuquerque, NM 87185, USA}

\date{\today}

\begin{abstract}
Compact cold-atom sensors depend on vacuum technology. One of the major limitations to miniaturizing these sensors are the active pumps---typically ion pumps---required to sustain the low pressure needed for laser cooling.  Although passively pumped chambers have been proposed as a solution to this problem, technical challenges have prevented successful operation at the levels needed for cold-atom experiments.  We present the first demonstration of a vacuum package successfully independent of ion pumps for more than a week; our vacuum package is capable of sustaining a cloud of cold atoms in a magneto-optical trap (MOT) for greater than \days\ days using only non-evaporable getters and a rubidium dispenser.  Measurements of the MOT lifetime indicate the package maintains a pressure of better than $2\times10^{-7}$ Torr.  This result will significantly impact the development of compact atomic sensors, including those sensitive to magnetic fields, where the absence of an ion pump will be advantageous.
\end{abstract}

\maketitle

\section{Introduction}
Atomic sensors, which make use of the inherent precision of atomic energy levels, are moving from the laboratory, where they have pushed the limits of precision measurements \cite{parker2018,rosi2014,brewer2019}, to applications in the field, where the requirements are shifted toward portability, reliability, and integration \cite{Bongs_2019}. Successful miniaturization of vacuum technology for cold-atom sensors will impact a wide range of  applications including gravimeters, accelerometers, gyroscopes, clocks, magnetometers, and gravity gradiometers.

These sensors must include an ultra-high vacuum (UHV) system capable of sustaining a cloud of cold atoms. This cloud is often produced using a magneto-optic trap (MOT). With existing UHV technology, both the pumping apparatus and the gauges require many cubic centimeters at best \cite{basu2016}, in addition to the chamber itself.  While passively pumped vacuum packages utilizing non-evaporable getters (NEGs) have been proposed \cite{Rushton:2014,dellis2016lowHe}, there has been doubt as to whether or not they are sustainable over the time scales needed for cold-atom sensors \cite{sebby2016honeywell,basu2016}.  Getters have been used in miniature ion-clock chambers \cite{gulati2018miniatured,jau2012low}, however these systems take advantage of a getter-pump's inability to pump noble gases, a severe disadvantage for applications involving MOTs. 

We present a vacuum chamber that uses passive pumping to maintain pressures sufficient for a MOT in excess of \days\ days. Using the MOT as a rough pressure gauge, we estimate that the system sustains a background pressure of better than $2\times10^{-7}$ Torr \cite{Arpornthip:2012, Moore_2015}.  
Although this pressure is relatively high for cold-atom systems, it is sufficient for applications such as high data rate atom interferometry \cite{Rakholia2014}. While efforts are being made by others to miniaturize vacuum systems for the purpose of atomic sensors \cite{mcgilligan2020, sebby2016honeywell,dellis2016lowHe}, to our knowledge this is the first demonstration of a passively pumped chamber sustaining a MOT for more than a week \cite{boudot2020}. 

\section{Vacuum Design and Fabrication}

\begin{figure}
\includegraphics[]{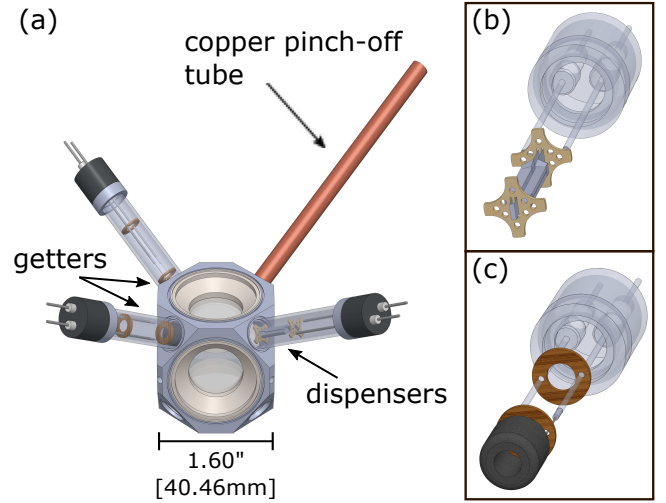}
\caption{\label{fig:pkg} (a) The vacuum package consists of a titanium body with six sapphire windows and other components laser-welded onto it.  (False transparency reveals the inside components of the arms, shown in the insets in more detail.) A copper pinch-off tube allows pumping down with standard turbo and ion pumps and has been successfully pinched off to create a cold-welded seal that preserves the vacuum.  (b) Two rubidium dispensers are held in place with alumina spacers. (c) Detail of one of the non-evaporable getters.}
\end{figure}

The primary challenge of the design is to make a chamber that can sustain a high level of vacuum without dependence on active pumping, such as an ion pump, while allowing the optical access needed for cold atom experiments.  We utilize non-evaporable getters, which passively pump the chamber by means of chemisorption\cite{SAESgetters}.

Since the NEGs do not pump rare gases, any helium permeation will limit the lifetime of the vacuum.  Although it is easy to find materials for the body of the chamber that will not allow helium permeation, finding a transparent material for the windows with this property is more challenging.  Aluminosilicate \cite{dellis2016lowHe} and alumina \cite{ohanlon2005,perkins1973permeation} perform better than borosilicate, however the latter presents some fabrication challenges. We utilize sapphire, which has no documented helium permeation that the authors are aware of, and use C-cut windows to minimize birefringence and maximize strength.  The body of the package is fabricated from commercially pure titanium since it matches well with the coefficient of thermal expansion of sapphire, has a much lower hydrogen outgassing rate than stainless steel\cite{takeda2011hydrogen}, and is nonmagnetic. 

The configuration of the vacuum package is shown in Figure \ref{fig:pkg}.  The sapphire windows (MPF Products, Inc.) braised into titanium frames are laser-welded into the electropolished titanium package body. Four corner tubes support operation: two with getters (SAES St172/HI/7.5-7), one with two dispensers (SAES Rb/NF/3.4/12FT10+10), and a copper tube which is connected to a standard UHV pumping apparatus prior during initial pump-down and testing, but is later pinched off to form a cold-welded seal.  The getters and dispensers are housed in commercially pure titanium tubing and are connected to the outside of the chamber via custom electrical feedthroughs, as shown in Figure \ref{fig:pkg}.  The external volume of the package is approximately 70 mL. 

The entire assembly is baked in a vacuum furnace at 400\degree C for seven days at $1\times10^{-7}$ Torr \cite{footnote}. During this time, the package interior volume is evacuated independently with total pressure and residual gas pressure monitoring to assure sufficient exhaust parameters are achieved and that unwanted gases from the getters and dispensers are removed as they are electrically activated. Final post-exhaust pressures of $3.2\times10^{-9}$ Torr are observed on the vacuum package near the turbo pump. Helium leak tests are performed throughout the fabrication and exhaust process to ensure vacuum integrity.  On the final package, the ion pump current of 0.7 nA gives an estimated pressure of $2.7 \times10^{-11}$ Torr. 

\section{Test Setup and Protocol}\label{sec:theory} 
\begin{figure}
\includegraphics[]{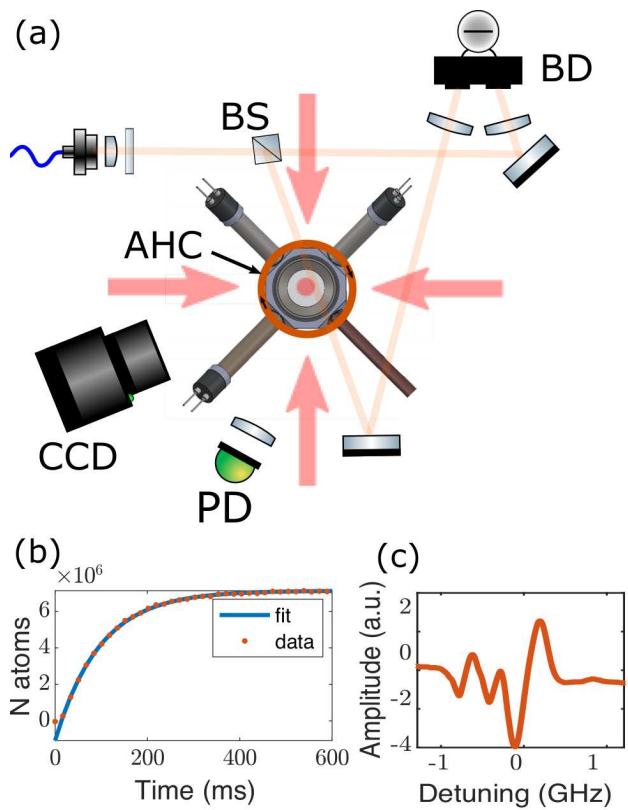}
\caption{\label{fig:exp} The vacuum package is tested using fluorescence from atoms in a MOT. (a) Six circularly polarized 780-nm laser beams (four of which are indicated by the large arrows) locked 8 MHz to the red of the F=2 to F’=3 line in $^{87}$Rb provide cooling.  A second laser tuned to the F=1 to F’=2 line is mixed in with the axial cooling beams and serves as a repump.  Anti-Helmholtz coils (AHC) are mounted directly on the chamber. We observe loading curves by switching the current to these coils off and then back on. The cloud of atoms is imaged on a CCD; load curves (b) are obtained from the CCD and photodiode (PD).  A 795 nm probe beam is split on a beamsplitter (BS), with one arm passing through the interior of the chamber, while the other serves as a reference. The difference measurement of the absorption probe is sent from the balance detector (BD) to a lock-in amplifier; the amplitude of the signal shown in (c) is used to calculate the rubidium pressure.}
\end{figure}

We utilize a MOT to characterize our vacuum package.  Not only does this demonstrate the viability of the chamber for cold-atom experiments, but it serves as an indication of the evolution of the background pressure over time.  

The trap loading dynamics are well described by \cite{Moore_2015, Arpornthip:2012}:
\begin{equation}\label{eq:dN}
\frac{dN}{dt}=R-\Gamma N,
\end{equation}
where $N$ is the number of atoms in the MOT, $R$ is the loading rate, proportional to the rubidium pressure $P_{Rb}$, and $\Gamma=\gamma_{Rb} P_{Rb} +\gamma_{bk} P_{bk}+\Gamma_0$ is the loss rate.  This loss rate depends on the rate $\gamma_{Rb}$ of collisions with rubidium, the rate $\gamma_{bk}$ of collisions with other background gasses of pressure $P_{bk}$, and a density-dependent factor $\Gamma_0$ accounting for two body collisions within the cold cloud. In a typical MOT, the density of the trapped atoms is limited by light scattering forces; in this constant density limit $\Gamma_0$ can be approximated as constant and one obtains an exponential loading curve \cite{Arpornthip:2012,Moore_2015}:
\begin{equation}\label{eq:N}
 N= N_0(1-e^{-t/\tau}),
\end{equation}
where $N_0=R\tau$ is the total number of atoms loaded with time constant $\tau=1/\Gamma$.  $N_0$ and $\tau$ are measured by observing the fluorescence of the atoms as a MOT is loaded \cite{Steck85}.  A representative loading curve and fit is shown in Figure \ref{fig:exp} (b), which was collected after switching on the anti-Helmholtz coils. The CCD counts are calibrated to determine the number of atoms \cite{Steck85}.

Arponthip, et al.\cite{Arpornthip:2012} showed that $\tau$ can be used to estimate the background pressure in typical UHV systems.  Specifically, the pressure inside the chamber is well approximated by 
\begin{equation}\label{eq:P_est}
P_{\text{vacuum}} <(2\times 10^{-8} \text{ Torr} \cdot \text{s})/\tau.
\end{equation}
Despite a few caveats \cite{Arpornthip:2012}, this gives a good upper bound on the background pressure of the vacuum system.  We also monitor the pressure of Rb in the vacuum package directly using a probe laser on the D1 line.

A schematic of the test setup is shown in Figure \ref{fig:exp}.  The 780 nm cooling laser is locked 8 MHz to the red of the F=2 to F’=3 resonance of $^{87}$Rb.  The repump light is resonant with the F=1 to F'=2 transition of  $^{87}$Rb.  Both lasers are coupled into polarization maintaining fibers and distributed to the vacuum test setup via splitters.
	
A CCD and a photodiode are used to measure loading curves. The majority of the data presented utilizes the results from the CCD; the photodiode serves to confirm particularly short loading times (Figure \ref{fig:dispchange}). The trapping magnetic field is generated with a pair of circular anti-Helmholtz coils which are switched via software control.  This method allows for background subtraction of florescence from atoms in the chamber, compared to those loaded into the trap.  

The density of rubidium in the chamber is measured via the absorption of another laser beam as it is swept through the D1 F=3 to F'=2,3 transition of $^{85}$Rb, for which the cross-section is calculated to be $9.2 \times 10^{-16} \text{m}^2$ \cite{Siddons2008}. From this density, a pressure is calculated.  We use a balanced detector to reduce the noise due to power fluctuations.  The signal to noise is further improved by use of a lock-in detector; the current of the laser is modulated at 100 kHz while it is swept across the resonance at 0.5 Hz.  The resulting amplitude of the dispersive lock-in signal shown in Figure \ref{fig:exp} (c) is proportional to the absorption of the probe. 

The vacuum package was initially set up with an ion pump. After establishing testing procedures, the ion pump was switched off.  Following successful operation with the pump off for a month, the ion pump was switched back on in preparation for pinching off the copper tube.  During the month of pumpless operation, the loading times and MOT atom number were around 100 ms and $7\times10^5$, respectively; prior to pinch-off, they were around 1 s and $7\times10^6$.  A pneumatic pinch-off tool (CPS HY-187-F) was used to pinch off the copper tube.  Results of the measured MOT loading parameters following pinch-off are shown in Figure \ref{fig:res1}.  

\section{Results}\label{sec:Results}

\begin{figure}
\includegraphics[]{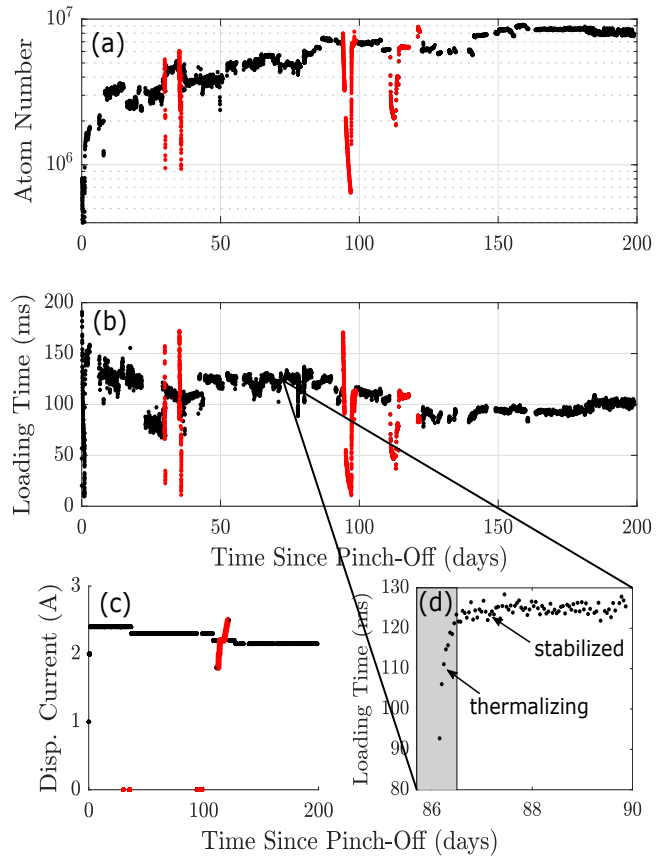}
    \caption{\label{fig:res1} Following pinch-off, (a) the number of atoms $N$ and (b) the characteristic loading times $\tau$ of a MOT in the passively pumped chamber are monitored over the course of \days\ days.  Other activity on the optical table required the laser enclosure curtains to be opened during the day, causing drifts like the one shown in (d), which shows the loading time over the course of a long weekend.  Many of these transients have been removed from (a) and (b) to better show the trends.  Major variations due to large changes in the rubidium dispenser current are highlighted in red, with the corresponding dispenser current used shown in (c). }
    \label{3figs}
\end{figure}

\begin{figure}
\includegraphics[]{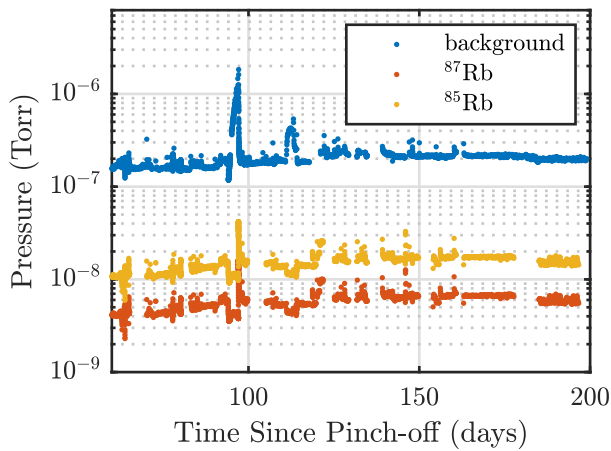}
\caption{\label{fig:pressure} The pressure of the background vapor in our vacuum package is estimated using Equation \ref{eq:P_est}, while the pressures of both isotopes of rubidium in the chamber are estimated using the absorption probe measurement.  Data is shown since calibration of this measurement on day 60.}  
\end{figure}

The primary result of this work is the demonstration that a passively-pumped vacuum chamber can support cold atom physics experiments for months.  Figures \ref{fig:res1} (a) and (b) show the number of atoms in the trap $N$ and the loading time constant $\tau$ over a period of \days\ days following pinch-off.  
Each data point is obtained by fitting Equation \ref{eq:N} to a measured loading curve, as in  \ref{fig:exp} (b). Experiments involving large changes of the dispenser current are highlighted in red.  The dispenser current is shown in \ref{fig:res1}(c).  Using the absorption probe and Equation \ref{eq:P_est}, we estimate the pressure of rubidium and the background gases in our chamber, as shown in Figure \ref{fig:pressure}.  After initial changes immediately following the pinch-off, the MOT loading parameters change relatively slowly, indicating that the passive pumping is able to maintain a vacuum in the system for many months.  Variation in the plots is due to a number of factors, which we divide into short-term transients and long-term trends.

The short day-to-day variability in MOT loading parameters is dominated by temperature changes caused by opening and closing the curtained enclosure around the optics table and other work going on in the lab.  This is highlighted in the example shown in Figure \ref{fig:res1} (d), which shows a data run taken over a long weekend; there is an initial thermalization period which we exclude from our analysis.  We attribute many of the short-term transients to a combination of temperature-caused misalignments and changes in the rubidium pressure due to ambient temperature.  To test the temperature dependance, we placed a temperature logger on the table next to the test setup. Temperature changes of a few tenths of a degree result in significant changes to both the atom number $N$ and the loading time constant $\tau$.  
These variations are consistent with the observed variations in the Rb pressure. More details can be found in the Supplemental Material \cite{sup}.

Long-term trends are both more difficult to explain and more interesting.  The variation in the average values in Figure \ref{fig:res1} (a-b) may be a result of hysteresis in the alignment in various parts of the test setup, as well as disturbances to the alignment due to other activity in the lab. The alignment of the system has been optimized several times to maximize the atom number.  These realignments tend to cause step changes in N, for example near day 140 in Figure \ref{3figs} (a). Our absorption measurement of the rubidium density indicates an increase in the amount of rubidium in the chamber over the \days\ days;  this likely plays a role in the increased MOT atom number and decreased loading time seen over this period.  For example, from day 60 to 180, the rubidium pressure increased by around $1.4\times10^{-8}$ Torr (Fig \ref{fig:pressure}); based on an estimated loss coefficient for Rb-Rb collisions \cite{Arpornthip:2012}, this could contribute to the MOT loss rate $\Gamma=1/\tau$ by 0.6 s$^{-1}$.  During the same time, $\tau$ decreases from 120 ms to 90 ms (Fig \ref{3figs}), corresponding to an increase $\Delta \Gamma \approx3$ s$^{-1}$ in the loss rate.  Thus while the increase in rubidium plays a role, there are likely other effects contributing to changes in $\tau$.  

\begin{figure}
\includegraphics[]{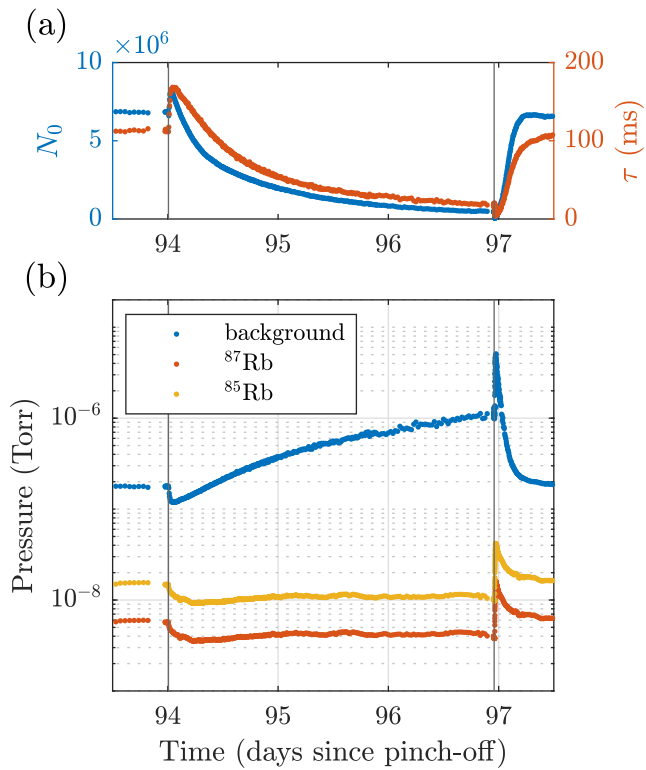}
\caption{\label{fig:dispchange} The rubidium dispenser is cycled off and on between days 93 and 98 at times marked with vertical lines.  (a) As expected, the number of atoms in the MOT drops, although this number trends toward a non-zero cloud size, indicating that the chamber could sustain operation without the dispensers for a significant period of time.  The loading time also decreases; this is used to estimate the rising background pressure (Eq. \ref{eq:P_est}) shown in (b), along with the rubidium pressures calculated from the absorption measurement.}
\end{figure}

In order to understand the role of the rubidium dispenser in our vacuum package, several dispenser current variation experiments were performed. The behavior of $N$ and $\tau$ during these experiments suggests that the Rb dispenser plays a significant role in maintaining the pressure. The result of one of these tests is shown in Figure \ref{fig:dispchange}. Naively one would expect from Equation \ref{eq:dN} that when the dispenser is turned off, the number of trapped atoms would begin to decrease as the amount of rubidium decreases, while the loading time would increase \cite{Moore_2015}. Surprisingly, the two parameters trend in the same direction; we hypothesize that this is indicative of a dispenser pumping effect---due to the gettering material of the dispenser, and possibly also due to the increased number of alkali atoms, which may serve to reduce the pressure of other gases in the chamber.  

This can be seen by plotting the background pressure estimated from loading curve fits and Equation \ref{eq:P_est} alongside the results of the absorption measurement, in Figure \ref{fig:dispchange} (b). When the dispenser is turned off (day 94), the increase in the background pressure appears to be a larger effect than the decrease in rubidium pressure. When the dispenser is turned back on, there is a spike in background pressure: this effect has been noted by others, and is likely due to a release of gas from both the chamber walls and the dispenser material as the dispenser temperature increases \cite{kohn2020clean}. After some time, the pressures return to their values prior to the cycling experiment.

While the dispenser pumping effects are significant, it is worth noting that this experiment also demonstrates that even after three days of leaving the dispenser off, the chamber still supports a cloud of around $5\times 10^5$ cold atoms.  

\section{Conclusion} 
We have demonstrated the successful design and testing of a portable vacuum package that can sustain vacuum levels low enough for cold atoms using only passive pumping via non-evaporable getters. The success of our design is undoubtedly due to a combination of factors, including the low helium-permeability of the materials and the fabrication procedures. We are in the process of building and testing other vacuum chambers with different parameters; the results of these future tests will shed light on which parts of the design and fabrication are most critical.  

Our design has a number of advantages for atomic sensors. While some have proposed the use of other non-magnetic ion pumping mechanisms \cite{sebby2016honeywell}, eliminating the need for an ion pump altogether presents a clear advantage. We have successfully driven 6.8 GHz microwave transitions in rubidium in the chamber, demonstrating that the metal body is not an obstacle to such atomic state manipulation. Finally, in contrast to chip-focused designs \cite{mcgilligan2020}, the six windows allow optical access for the counter-propagating beams typically used in atom interferometer applications. 

The behavior of the MOT loading time and atom number in response to different changes, such as the current in the dispenser or the temperature of the environment will be the subject of future study. In such a small chamber, subtleties arising from the likely pumping of the dispenser present both challenges and advantages that we would like to understand better. The dispenser requires 2.15 A current to maintain optimal conditions, or 2.22 W of power. Continued investigation is required to develop techniques to achieve substantially reduced power consumption while maintaining the appropriate alkali \cite{kang2017low,kohn2020clean} and background pressures.

Our result represents significant progress toward reducing the size, weight, and power consumption of atomic sensors. We expect that it will drive the development of more robust sensors which can be used in a wide range of applications, from fundamental physics such as measurements of gravity in space, to more application-focused developments in civil engineering. It is particularly well-suited for inertial sensing devices, where the demand is high for a robust and compact vacuum package.

\section*{Acknowledgments}
The authors would like to Melissa Revelle for her contributions to the testing, and members of John Kitching's group at NIST for their ideas and feedback. 
This work was supported by the Laboratory Directed Research and Development program at Sandia National Laboratories. Sandia National Laboratories is a multimission laboratory managed and operated by National Technology $\&$ Engineering Solutions of Sandia, LLC, a wholly owned subsidiary of Honeywell International Inc., for the U.S. Department of Energy’s National Nuclear Security Administration under contract DE-NA0003525.  This paper describes objective technical results and analysis. Any subjective views or opinions that might be expressed in the paper do not necessarily represent the views of the U.S. Department of Energy or the United States Government.

\section*{Author's Contributions}
G. W. Hoth and B. J. Little contributed equally to this work.


%


\bibliography{vacuumpaper}

\begin{thebibliography}{25}
\bibitem [{foo()}]{footnote}%
  \BibitemOpen
  \href@noop {} {\bibinfo {title} {We are uncertain that baking the miniature
  vacuum within a larger vacuum furnace is necessary. {We} did this to ensure
  there was not helium permeated into any of the materials, but none of the
  materials should allow any helium permeation so this step may have been
  unnecessary}}\BibitemShut {NoStop}%
\end{thebibliography}

\end{document}